\documentclass[conference]{IEEEtran}
\usepackage{amsmath,amsfonts,amssymb}
\usepackage{algorithmic}
\usepackage{algorithm}
\usepackage{array}
\usepackage[caption=false,font=normalsize,labelfont=sf,textfont=sf]{subfig}
\usepackage{textcomp}
\usepackage{stfloats}
\usepackage{graphicx}
\usepackage{cite}
\usepackage{booktabs}
\usepackage{soul,xcolor}
\usepackage{xurl}
\usepackage{makecell}
\usepackage{adjustbox}
\usepackage{ragged2e}
\usepackage{tabularx}
\usepackage{multirow}
\usepackage{verbatim}
\usepackage[colorlinks=true, urlcolor=black, linkcolor=black, citecolor=black, breaklinks=true]{hyperref}

\usepackage{threeparttable}
\usepackage{tikz}
\usepackage{tikz-timing}
\usepackage[percent]{overpic}

\begin{document}

\title{A High-Throughput FPGA Architecture for Real-Time TCP-SYN Scan Detection}

\author{Faisal Saeed\IEEEauthorrefmark{1}, Mohammad Fahad\IEEEauthorrefmark{1}, Ayesha Javaid\IEEEauthorrefmark{1}, Christian Doerr\IEEEauthorrefmark{4} and Muhammad Ali Siddiqi\IEEEauthorrefmark{1}\\
\IEEEauthorrefmark{1}Department of Electrical Engineering, Lahore University of Management Sciences, Lahore, Pakistan \\
\IEEEauthorrefmark{4}Hasso Plattner Institute, Potsdam, Germany
}

\maketitle

\begin{abstract}
TCP-SYN port scanning often precedes cyber-attacks, and early detection of scanner fingerprints embedded in packet headers can provide timely intrusion alerts. Existing approaches are either too computationally expensive for line-rate operation or limited to offline analysis. This brief presents a lightweight FPGA architecture for reconfigurable line-rate fingerprint detection, where each fingerprint is compiled into a shallow Boolean LUT tree, enabling parallel evaluation with constant two-cycle latency regardless of fingerprint count, while resource cost grows linearly with fingerprint count. This detection core is decoupled from a MAC-layer frontend that performs streaming field extraction with no frame buffering or higher-layer state, allowing deployment across different line rates by modifying only the frontend. A Python framework automatically compiles Boolean expressions into synthesizable HDL, eliminating manual RTL changes. For TCP-SYN port-scan fingerprint detection, the architecture uses approximately 0.5\% LUTs at 10\,Gbps on a Versal VCK190 for 18 deployed fingerprints, with capacity for over 2,000 concurrent fingerprints, and under 2.5\% on a Virtex-6 at 1\,Gbps, with a detection latency of 10\,ns at both rates, three to four orders of magnitude below typical per-packet processing latency in software intrusion-detection systems. The system was cross-validated against a software re-implementation on an 8-hour production packet trace, confirming detection correctness with zero false positives/negatives.
\end{abstract}


\section{Introduction}
\IEEEPARstart{P}{ort} scanning is a common precursor to cyber-attacks, used to identify vulnerable services. TCP-SYN scanning predominates because SYN packets require no session state, enabling rapid sweeps across large address ranges. Detecting such activity early can reveal malicious intent before an attack advances, but at contemporary data rates detection must operate at line speed (wire rate) to remain effective. Modern scanners such as Masscan and ZMap limit probes per destination and randomize target order, rendering behavior-based detection ineffective~\cite{staniford,Griff123}; anomaly detection cannot isolate such sparse scanning from legitimate SYN traffic; and payload inspection is inapplicable since SYN packets carry no payload. Among detection signals, \textit{fingerprints} are the most actionable: scanners embed deterministic relationships in header fields, e.g.\ \texttt{tcp\_seq}\,$=$\,\texttt{ip\_dst} in Masscan~\cite{masscan}, a fixed \texttt{tcp\_window} in ZMap~\cite{zmap}, because the formula lets attackers identify returning ACKs without per-destination bookkeeping, making such patterns an unavoidable artifact of stateless scanning. This motivates a fingerprint-based detection approach realized directly in hardware.

Because each fingerprint is a fixed Boolean function of a bounded set of header fields, it compiles naturally onto a shallow combinational LUT (Lookup Table) tree. We exploit this to design an FPGA architecture in which every deployed fingerprint evaluates in parallel, so detection latency is independent of how many fingerprints are active, while resource cost grows only linearly with fingerprint count, in contrast to a software rule engine, which scales linearly in per-packet processing time as fingerprints are added. Unlike an ASIC, the FPGA implementation can also be reconfigured as scanners evolve their fingerprints to evade detection, without a fabrication cycle.  

\begin{figure}
    \centering
    \includegraphics[width=0.9\linewidth]{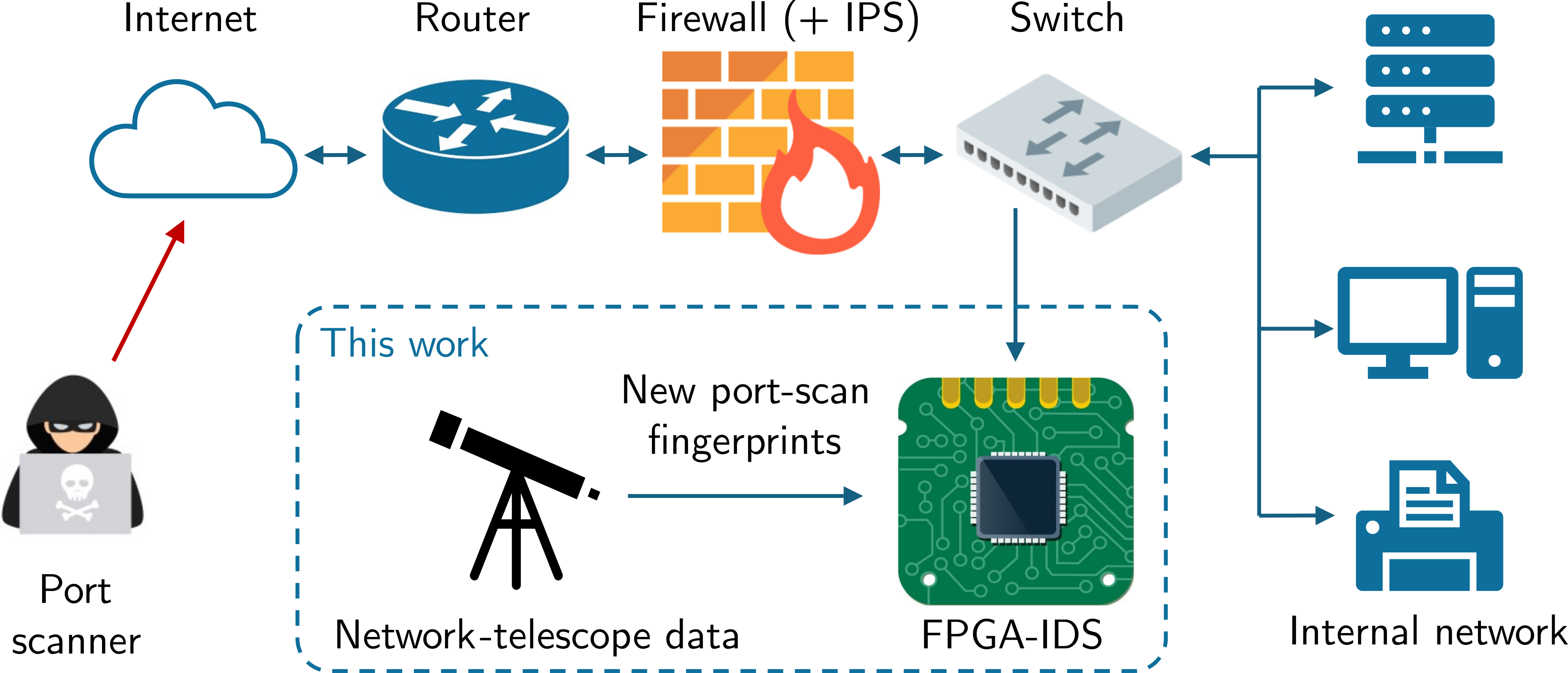}
    \caption{Deployment of the proposed FPGA-IDS within a network topology.}
    \label{fig:overview_fpga_ids}
\end{figure}

This work makes the following contributions.
(i) We present an FPGA architecture for fingerprint-specific TCP-SYN scan detection whose detection latency ($T_\mathrm{det}$) is independent of fingerprint count, operating entirely at the MAC interface and completing detection before the next frame's header arrives, at both 1\,Gbps and 10\,Gbps.
(ii) We develop a Python-based RTL generation framework with a defined Boolean fingerprint grammar that produces synthesizable HDL without manual RTL modification. (iii) We demonstrate architectural modularity by reusing a common fingerprint detector unchanged across XGMII (10 Gbps) and GMII (1 Gbps) frontends, modifying only the interface-adaptation layer. We implement the 10 Gbps variant on a Versal VCK190, and the 1 Gbps variant on Virtex-6.
\figurename{~\ref{fig:overview_fpga_ids}} summarizes the deployment.

\section{Related Work}
\textbf{Scan detection in software.} Behaviour-based detectors flag scanners using per-source thresholds or failure counts~\cite{Yegneswaran}, but modern scanners evade this by spreading probes thinly across many sources~\cite{staniford}. Even a hardware implementation of Threshold Random Walk (TRW)-style containment~\cite{weaver2004containment} would still need to track connection state per source, something a bufferless, line-rate pipeline cannot maintain. Fingerprint-based approaches instead recover deterministic relationships between header fields, but only offline: Griffioen et al.~\cite{Griff123} use graph clustering over data from a network telescope\footnote{A network telescope passively monitors a large block of unused IP addresses, capturing only unsolicited traffic such as scans and backscatter.}, and Tanaka et al.~\cite{Tanaka1,Tanaka2} use genetic search on the same type of data. Both rely on traffic recorded from unused address ranges, where scan traffic is easy to isolate; on real networks, the same scan traffic is mixed with ordinary flows and far harder to separate. 

\textbf{FPGA-based approaches} fall into two categories: payload matching and header matching. Payload-inspection engines match Snort-style byte patterns at high speed but consume large amounts of memory: classic string-matching circuits require 4--5 logic cells per pattern character~\cite{sourdis2003}, Kim et al.~\cite{kim} require one on-chip memory block per pattern (capped at 265 rules, 1.06\,Gbps), and Pigasus~\cite{pigasus} reaches 100\,Gbps only using SmartNIC-class FPGA resources for packet reassembly and multi-pattern matching. Header-only detectors use fewer resources but are either slow or not fingerprint-specific: Dakhil et al.~\cite{9800025} match stored rules using a state machine at 100\,Mbps, Kang et al.~\cite{8969630} build a whitelist of 7 expected header fields at 100\,Mbps, and Das et al.~\cite{das} and Farooq et al.~\cite{raw} target general anomaly or neural-network classification rather than explicit fingerprints. P4-to-FPGA compilers~\cite{p4fpga} and platforms such as Corundum~\cite{corundum} also parse headers at 10--100\,Gbps, but target general-purpose packet processing using tens of thousands of logic cells. Unlike prior work, our detector is fingerprint-specific, online, and line-rate, with no content-addressable memory, per-rule storage, or buffering, and can be added as a pre-filter stage within any of these pipelines. Table~\ref{tab:litreview} summarizes the landscape.

\begin{table}[!t]
\centering
\caption{Overview of the State of the Art}
\label{tab:litreview}
\begin{threeparttable}
\begin{adjustbox}{max width=\columnwidth}
\renewcommand{\arraystretch}{0.95}
\setlength{\tabcolsep}{3pt}
\begin{tabular}{@{}ccll@{}}
\hline
\textbf{Ref} & \textbf{T}$^{\dagger}$ & \textbf{Approach} & \textbf{Limitation} \\
\hline
\cite{Yegneswaran}      & B & Dst port/IP thresholds   & $\geq$5 scanners/hr; offline \\
\cite{weaver2004containment} & B &  TRW containment & Per-source state \\
\cite{Griff123}        & F & Graph clustering, XOR/shift & $\leq$131K comparisons; offline \\
\cite{Tanaka1,Tanaka2}  & F & Genetic-evolved fingerprints  & Labeled data; offline \\
\cite{das}              & H & PCA anomaly detection    & Not fingerprint-specific \\
\cite{9800025}          & H & FSM match vs.\ ROM       & $\leq$100\,Mbps; ROM-bound \\
\cite{8969630}          & H & Whitelist 7-tuple matching       & 100\,Mbps prototype \\
\cite{raw}              & H & LUT-mapped NN, raw bytes & Not fingerprint-specific \\
\cite{kim}              & H & Shift-And, IEC\,61850    & $\leq$265 rules; 1 BRAM/rule \\
\cite{pigasus}          & H & FPGA-first Snort DPI     & Payload; SmartNIC-class cost\\
\hline
\end{tabular}
\end{adjustbox}
\begin{tablenotes}
\footnotesize
\item[$^{\dagger}$] Type --- B: behaviour, F: fingerprint, H: FPGA/hardware.
\item[] DPI: deep packet inspection; NN: neural network.
\end{tablenotes}
\end{threeparttable}
\end{table}

\section{Proposed Method}
\label{sec:proposed-method}
We propose an FPGA architecture that detects fingerprints in TCP-SYN port-scanning traffic in real time, co-designed around one constraint: detection must complete within the inter-frame interval using only TCP-SYN header fields, with no frame buffering or higher-layer state. This splits the design into a \emph{detection core}: fingerprint detector and TCP-SYN validation, invariant across line rates, and an \emph{interchangeable frontend}: only the MAC, which recovers byte-aligned header fields from the Physical Coding Sublayer (PCS), changes with the interface, spanning the MII (Media Independent Interface) family, XGMII (10\,Gbps), GMII (1\,Gbps), and MII/RMII (10/100\,Mbps). The separation drives three choices: the MAC acts as a \emph{streaming field extractor}, registering header fields into named slots as bytes arrive and eliminating BRAM storage and post-reception latency; each signature uses a \emph{uniform two-stage pipeline}, the maximum depth that still completes within the inter-frame interval for minimum-size frames; and the Python generator (Fig.~\ref{fig:architecture}A) \emph{separates the TCP-SYN parsing logic from a reconfigurable fingerprint array}, so new signatures can be regenerated without modifying the surrounding detection core. The complete RTL and framework are available~\cite{anonymous2026code}.

\subsection{Baseline Frontend: GMII and Below (1\,Gbps and Slower)}
Interfaces at 1\,Gbps and below are variable-speed and, below 1\,Gbps, sub-byte-wide, requiring a link-speed detector and an accumulator to normalize incoming data before it reaches the shared detection core (Fig.~\ref{fig:architecture}B).

\subsubsection{Link-Speed Detector}
To distinguish 10/100/1000\,Mbps, a time-based frequency method is used (Fig.~\ref{fig:architecture}D). The unknown receive clock drives a slow counter while a stable 200\,MHz reference increments a 16-bit reference counter; on saturation the slower counter is sampled and crossed via a 2-FF synchronizer~\cite{cdc}. The observation window is $\frac{2^{16}}{200\times10^{6}} \approx 327.68\,\mu\text{s}$.

\subsubsection{Accumulator}
Because sub-byte interfaces deliver narrow words (4-bit MII nibbles, RMII di-bits), an accumulator normalizes incoming data to byte alignment (Fig.~\ref{fig:architecture}C), aligning successive words before the FIFO and managing control signals. At 1\,Gbps the byte-wide GMII data passes through with alignment already satisfied.

\subsubsection{Asynchronous FIFO}
The FIFO crosses between the 200\,MHz system clock and a 125\,MHz/25\,MHz/2.5\,MHz receive-clock domain, using the Gray-coded pointer design of Cummings and Alfke~\cite{asyncfifo}. With read enable permanently asserted, the FIFO stays near-empty and never buffers a complete frame, consistent with the streaming design philosophy. Writes scale with line rate: once per two nibbles at 10/100\,Mbps, every cycle at 1\,Gbps.

\subsubsection{Packet Processor}
The processor parses byte-serially, detecting the preamble (\texttt{0x55}) and SFD (Start Frame Delimiter, \texttt{0xD5}) with 64 bit-times of margin (Fig.~\ref{fig:architecture}E). Header fields are extracted after detection completes. A valid TCP-SYN packet must satisfy EtherType\,$=$\,\texttt{0x0800}, IPv4 version\,$=$\,4, Internet Header Length (IHL)\,$=$\,5, Protocol\,$=$\,6, SYN set, Data Offset (DO)\,$=$\,5, and valid checksums; these checks are evaluated combinationally on registered values, and only compliant packets are forwarded. Restricting to IHL\,$=$\,5/DO\,$=$\,5 removes the need for variable-length parsing, matching how stateless scanners such as Masscan and ZMap operate by default~\cite{masscan,zmap}. No payload beyond byte 53 is read, stored, or processed. At 1\,Gbps the 96\,ns inter-frame gap (IFG) provides 19 cycles at 200\,MHz, sufficient for all verification and matching.

\subsubsection{MAC State Machine}
After reset, the state machine waits for the link to become active. Once active, it triggers the Link Speed Detector and holds submodules in reset until the speed is determined (327.68\,$\mu$s). Once the link speed is known, it activates all submodules.

\subsection{10\,Gbps XGMII: A Fixed-Speed Simplification}
XGMII removes the Link-Speed Detector since it is fixed-speed and byte-aligned: the receive clock runs at a single 156.25\,MHz with no auto-negotiation, and each cycle delivers 64 bits (8 bytes) already byte-aligned.

\subsubsection{Accumulator}
Although data arrives byte-aligned, residual intra-word alignment is still required, since the Start control character can begin at byte 0 or byte 4 within the 64-bit word. The accumulator removes idle characters from the received block and aligns the Start control character with the rest of the preamble when it begins at byte 4.

\begin{figure*}[!t]
\centering
\includegraphics[width=\linewidth]{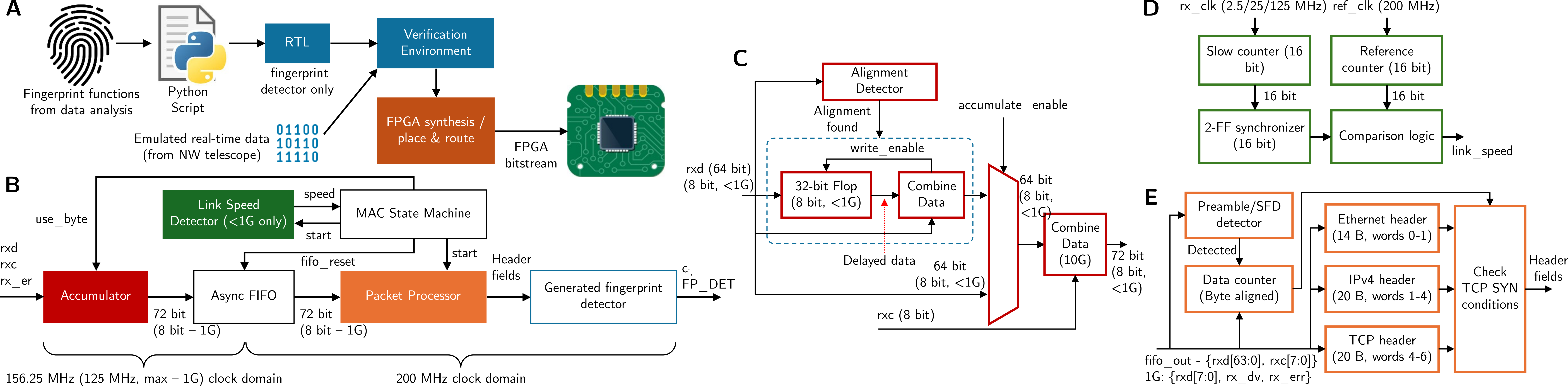}
\caption{A: Proposed tool flow for detection of fingerprints. B: System architecture. C: Accumulator. D: Link Speed Detector. E: Packet processor.}
\label{fig:architecture}
\end{figure*}

\subsubsection{Asynchronous FIFO}
The FIFO widens to 72 bits (64 data + 8 control) and crosses between the 156.25\,MHz receive clock and the 200\,MHz system clock.

\subsubsection{Packet Processor and MAC State Machine}
The packet processor consumes 64-bit words at once rather than byte-serially (Fig.~\ref{fig:architecture}E), locating the preamble/SFD pattern within a single word and extracting header fields from fixed word positions: Ethernet (14\,B) in words 0--1, IPv4 (20\,B) in words 1--4, TCP (20\,B) in words 4--6.
The state machine reduces to two states, \texttt{IDLE} and \texttt{ENABLE}, releasing submodules from reset on link-up with no 327.68\,$\mu$s wait.

\subsection{Fingerprint Detector and RTL Generator}
Each fingerprint is a two-stage pipeline: one cycle to register the header vector $\mathbf{h}$, one to evaluate $\mathrm{match}_i=f_i(\mathbf{h})$. A top-level module asserts $\mathtt{FP\_DET}=\bigvee_{i=1}^{N}\mathrm{match}_i$ with a 16-bit counter per signature. Since the detector RTL is reused unchanged, only the frontend differs between interfaces; each interface is still synthesized to its own bitstream, since the frontend and detector are integrated into a single top-level design.
A fingerprint reduces its header relations to one match bit ($\mathrm{match}_i$), realised as a shallow tree of 6-input LUTs, identical across Virtex-6 and Versal. An $n$-bit equality against a constant, where $n$ is the combined width of the header fields involved, uses $\lceil n/6\rceil$ comparison LUTs plus a reduction tree: \texttt{tcp\_window==14600} ($n{=}16$) needs 3 comparison + 1 reduction = 4~LUTs. A two-field relation such as \texttt{ip\_dst}$\oplus$\texttt{tcp\_seq==0} ($n{=}32$) folds 3 bit-pair equality checks into each LUT6's single output, giving $\lceil 32/3\rceil=11$ comparison LUTs and a two-level reduction ($\lceil 11/6 \rceil + 1 = 3$~LUTs), 14~LUTs total.
These are analytic upper bounds; synthesis often does better via LUT combining, and the measured worst case is 71~LUTs / 130~registers (Virtex-6) and 26~LUTs / 130~registers (Versal). Because all $N$ submodules evaluate in parallel, $T_\mathrm{det}=2$~cycles regardless of $N$, while LUT area grows linearly in $N$.

The \textit{RTL generator} accepts a restricted Boolean expression language over named header fields, designed so that every expression reduces to a single combinational match bit evaluable within one clock cycle:

\begin{scriptsize}
\begin{verbatim}
fingerprint := relation {('&&'|'||') relation}
relation    := term ('=='|'!='|'<'|'<='|'>'|'>=') term
term        := factor {('&'|'|'|'^') factor}
factor      := field | field'['msb':'lsb']' | const
             | factor('<<'|'>>')const | '~'factor 
             | '('term')'
field       := ip_src | ip_dst | ip_id | ip_ttl 
             | ip_proto | tcp_sport | tcp_dport | tcp_seq 
             | tcp_ack | tcp_window | tcp_flags
const       := decimal | '0x' hex
\end{verbatim}
\end{scriptsize}

\begin{figure}[!t]
\centering
\includegraphics[width=1\linewidth]{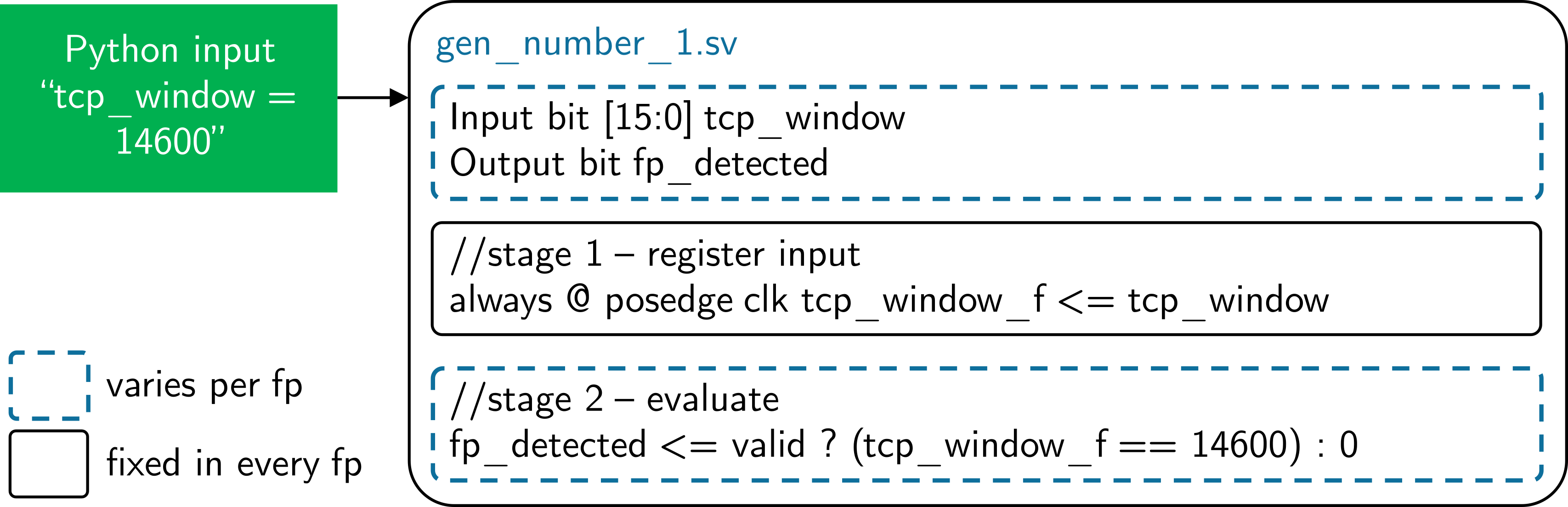}
\caption{Automated RTL generation. A one-line Python expression yields a complete two-stage submodule; only the highlighted regions vary.}
\label{fig:rtl}
\end{figure}

\noindent Bitwise operators ($\&$, $|$, $\oplus$, $\sim$) and constant shifts ($\ll$, $\gg$) build the compared value; shifts and masks are free rewiring or single-LUT operations. Comparison operators support equality, inequality, and magnitude tests, the last synthesizing to combinational magnitude comparators. Relations combine with logical \texttt{\&\&}/\texttt{||}. Multiplication and division are excluded, since they would break single-cycle evaluation. The framework (Fig.~\ref{fig:rtl}) emits one submodule per expression from a fixed template, varying only the port list and $f_i$; updates require regenerating HDL and reprogramming the bitstream.

\subsection{Threat Model}
\label{sec:threat_model}
Assuming header fields are otherwise random, a benign packet matches a $k$-bit fingerprint with probability $2^{-k}$ (e.g.\ $2^{-32}$ for \texttt{tcp\_seq}\,$=$\,\texttt{ip\_dst}); across $N{=}18$ fingerprints, the combined chance of any accidental match is at most $N/2^{k_{\min}} = 18/2^{16} \approx 2.7\times10^{-4}$, with $k_{\min}{=}16$ the narrowest field. This bound is conservative rather than representative: real SYN traffic is not field-random, clustering around common OS/stack defaults. Fingerprint reliability is not re-derived here; Tanaka et al.~\cite{Tanaka1,Tanaka2} validate this set independently against recorded scan traffic. Evading fingerprint $i$ requires breaking an operationally required relationship (e.g.\ stateless sequence seeding), which alters the scanner's observable pattern and typically produces a new, detectable fingerprint that can be compiled to a bitstream in minutes once identified; full randomization evades all fixed signatures but forfeits the stateless generation that makes high-speed scanning viable.

\section{Implementation and Results}
\label{sec:results}
Eighteen well-known scanner fingerprints were deployed, including signatures from Masscan and ZMap and others validated in~\cite{Tanaka1,Tanaka2}. The 10\,Gbps XGMII design, including portability and scaling tests, was implemented on a Versal VCK190, while the 1\,Gbps GMII design was implemented on a Virtex-6 (XC6VLX240T). A UART logs flagged packets to a host PC for evaluation only (not part of the deployment pipeline); on a match, the 54-byte header plus 3 bytes of metadata are streamed and parsed into a \texttt{.pcap}.

\subsection{Timing Budget by Line Rate}
\begin{figure}[!t]
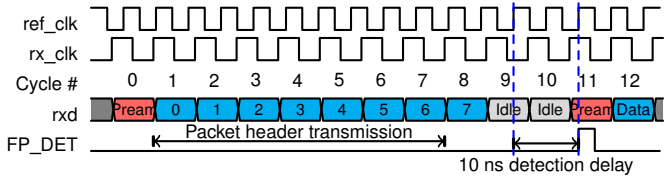


\begin{tikztimingtable}[
    timing/xunit=0.43mm,
    timing/dslope=0.1,
    timing/slope = 20,
    timing/rowdist=4mm,
    timing/name/.style={font=\sffamily\scriptsize}
]
ref\_clk   & 5H 5C 5C 5C 5C 5C 5C 5C 5C 5C 5C 5C 5C 5C 5C 5C 5C 5C 5C 5C 5C 5C 5C 5C 5C 5C 5C 5C 5C 5C 5C 5C 5C 5C 5C\\
rx\_clk  & 6.4C 6.4C 6.4C 6.4C 6.4C 6.4C 6.4C 6.4C 6.4C 6.4C 6.4C 6.4C 6.4C 6.4C 6.4C 6.4C 6.4C 6.4C 6.4C 6.4C 6.4C 6.4C 6.4C 6.4C 6.4C 6.4C 6.4C 3.2C\\
Cycle \#\\
rxd & 6.4U [fill=red!60]12.8D{Pream} [fill=cyan]12.8D{0} 12.8D{1} 12.8D{2}  12.8D{3}  12.8D{4}  12.8D{5}  12.8D{6}  12.8D{7}  [fill=black!15]12.8D{Idle} [fill=black!15]12.8D{Idle} [fill=red!60]12.8D{Pream} [fill=cyan]12.8D{Data} [fill=black!45]3.2U\\
FP\_DET & 5L  5L 5L 5L 5L 5L 5L 5L 5L 5L 5L 5L 5L 5L 5L 5L 5L 5L 5L 5L 5L 5L 5L 5L 5L 5L 5L 5L 5L 5L 5C 5C 5L 5L 5L \\
\extracode
\draw[thick, dashed, blue]
    (150,1.2) -- (150,-5.8);

\draw[thick, dashed, blue]
    (130,1.2) -- (130,-5.8);

\draw[very thick, black]
    (130,-4.9) -- (130,-5.3);

\draw[very thick, black]
    (150,-4.9) -- (150,-5.3);

\draw[thick, black, <->]
    (130,-5.1) -- (150,-5.1);

\node[black, below, font=\scriptsize] at (140,-5.3)
    {10 ns detection delay};

\node[black, above, font=\scriptsize] at (12.8,-3)
    {0};

\node[black, above, font=\scriptsize] at (25.2,-3)
    {1};

\node[black, above, font=\scriptsize] at (38,-3)
    {2};

\node[black, above, font=\scriptsize] at (50.8,-3)
    {3};

\node[black, above, font=\scriptsize] at (63.6,-3)
    {4};

\node[black, above, font=\scriptsize] at (76.4,-3)
    {5};

\node[black, above, font=\scriptsize] at (89.2,-3)
    {6};

\node[black, above, font=\scriptsize] at (102,-3)
    {7};

\node[black, above, font=\scriptsize] at (114.8,-3)
    {8};

\node[black, above, font=\scriptsize] at (127.6,-3)
    {9};

\node[black, above, font=\scriptsize] at (140.2,-3)
    {10};

\node[black, above, font=\scriptsize] at (153,-3)
    {11};

\node[black, above, font=\scriptsize] at (165.8,-3)
    {12};

\draw[very thick, black]
    (19.2,-4.8) -- (19.2,-5.2);

\draw[very thick, black]
    (108.8,-4.8) -- (108.8,-5.2);

\draw[thick, black, <->]
    (19.2,-5) -- (108.8,-5);

\node[black, above, font=\scriptsize] at (64,-5.3)
    {Packet header transmission};

\end{tikztimingtable}
\caption{Timing Diagram for Fingerprint Detection at 10 Gbps (XGMII) 
}
\label{fig:timing_diagram}
\end{figure}

The two-cycle detector asserts \texttt{FP\_DET} a fixed two cycles (10 ns at 200 MHz) after the final TCP header byte is registered. The window available for this differs by line rate. At 10 Gbps the IFG shrinks to 9.6 ns, under two 200 MHz cycles, so the two-cycle detector does not complete within the IFG in isolation; instead, because the 54-byte header is registered several words before a minimum frame ends, the preamble, frame tail, and IFG together provide the slack for minimum-size frames, which represent the worst case. The header is complete by cycle~7 and \texttt{FP\_DET} asserts at cycle~11, before the next frame's first header word arrives at cycle~12 (Fig.~\ref{fig:timing_diagram}). At 1 Gbps the 96 ns IFG spans 19 cycles at 200\,MHz, so detection completes strictly within the IFG.

\subsection{Resource \& Power Utilization}
\begin{figure}[!t]
\centering
\includegraphics[width=0.8\linewidth]{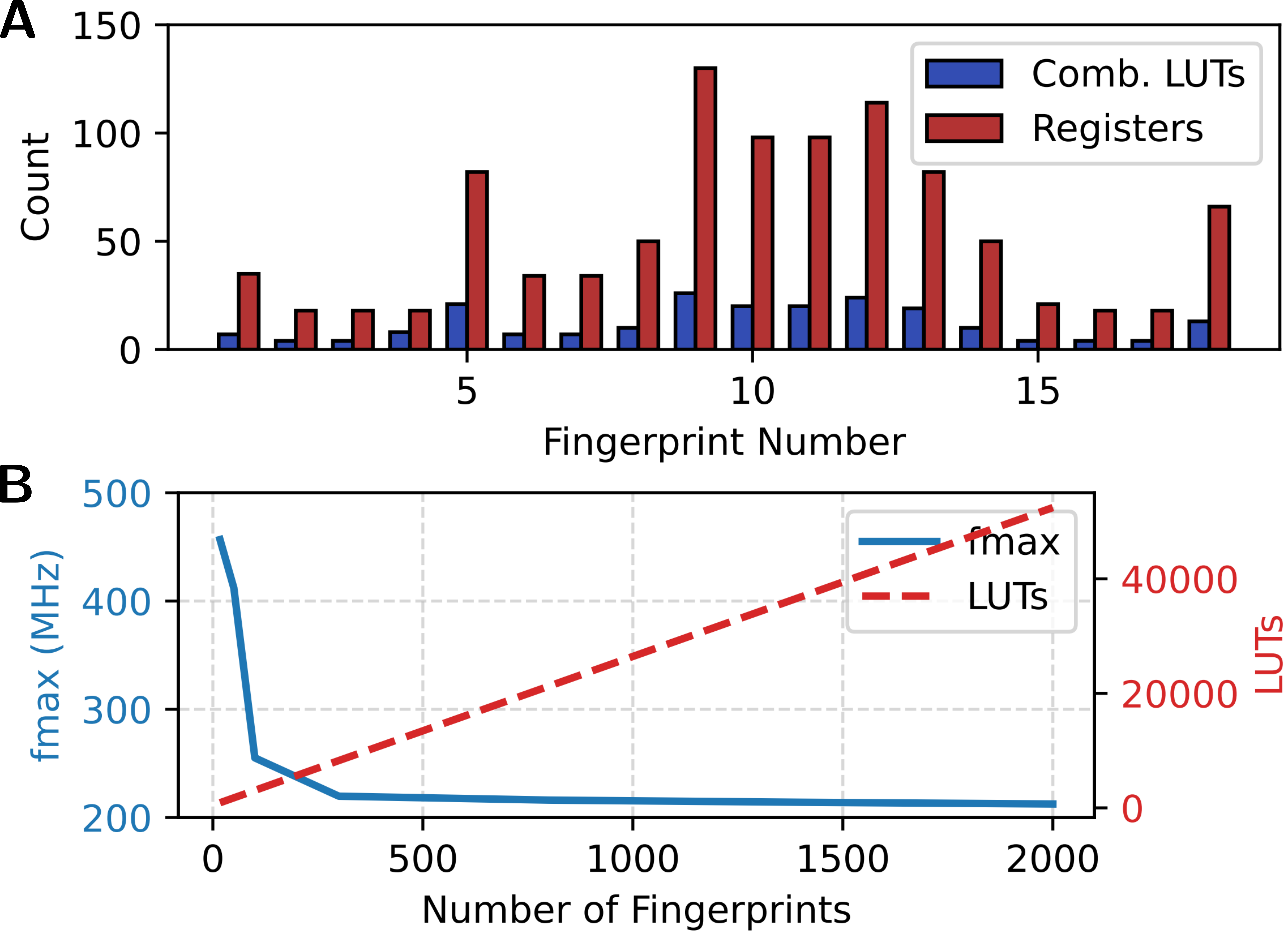}
\caption{Scalability of the proposed architecture. A: Resource consumption per fingerprint. B: $f_{\max}$ and LUT utilization versus fingerprint count.}
\label{fig:scalability}
\end{figure}

Table~\ref{tab:res} reports post-implementation utilization (excluding the evaluation UART).
Table~\ref{tab:hw_comparison} compares resource usage against prior FPGA-based detectors.
The 10\,Gbps XGMII variant on a Versal VCK190 uses 4{,}513 LUTs (0.50\%), 8{,}873 FFs (0.49\%), and 134 LUTRAM cells (clock-domain-crossing FIFO only). \figurename{~\ref{fig:scalability}A} shows the resource utilization by each fingerprint. Post-route primitive counts are consistent with the mapping: LUT6 dominates the detector logic, with carry chains (LUTCY) attributable to magnitude comparators and per-signature match counters. On the Virtex-6, the deployed design uses 1{,}595 LUTs (1.06\%) and 3{,}024 FFs, with the detector alone at 808 LUTs (0.54\%) and no DSP or BRAM. Post-route vectorless power analysis attributes 16 mW of dynamic power to the detection logic itself (7 mW fingerprint detector, 9 mW MAC frontend), with the complete 10 Gbps datapath including PCS/PMA, transceiver, and clocking at approximately 0.34 W.

\begin{table}[!t]
\centering
\begin{adjustbox}{max width=\columnwidth}
\begin{threeparttable}
\caption{Resource Utilization (UART Excluded, PCS/PMA\tnote{1} Included)}
\label{tab:res}
\begin{tabular}{@{}lrrrr@{}}
\toprule
\textbf{Module} & \textbf{LUT} & \textbf{FF} & \textbf{LUTRAM} & \textbf{$f_{\max}$} \\
\midrule
\multicolumn{5}{@{}l}{\textit{\textbf{Versal VCK190 (10G XGMII)}}}\\
Fingerprint Det. & 518 & 1273 & 0 & \\
MAC & 830 & 1382 & 86 & \\
PCS/PMA\tnote{1} & 3165 & 6218 & 48 \\
Total & 4513 (0.50\%) & 8873 (0.49\%) & 134 & 385\,MHz \\
\midrule
\multicolumn{5}{@{}l}{\textit{\textbf{Virtex-6 (1G GMII)}}}\\
Fingerprint Det. & 808 & 1349 & 0 & \\
MAC & 787 & 1675 & 23 & \\
Total & 1595 (1.06\%) & 3024 (1.00\%) & 23 & 357\,MHz \\
\bottomrule
\midrule
\multicolumn{5}{@{}l}{\textbf{Dynamic Power (post-route, vectorless)}}\\
\multicolumn{4}{@{}l}{\textit{Versal: detection logic (det.\ + MAC)}} & 16\,mW\\
\multicolumn{4}{@{}l}{\textit{Versal: full datapath\tnote{2}}} & 0.34\,W\\
\multicolumn{4}{@{}l}{\textit{Virtex-6: full design}} & 0.734\,W \\
\bottomrule
\end{tabular}

\begin{tablenotes}
\footnotesize
\item[1] Xilinx 10G Ethernet PCS$/$PMA (Physical Medium Attachment), responsible for interfacing with MAC, was included in utilization results as it is necessary for hardware implementation.
\item[2] Detector, MAC, PCS/PMA, transceiver, and clocking; excludes the Versal processing system of the evaluation platform (not part of the det. pipeline).
\end{tablenotes}
\end{threeparttable}
\end{adjustbox}
\end{table}

\begin{table}[!t]
    \centering
    \caption{Comparison with Prior Architectures}
    \label{tab:hw_comparison}
    \begin{adjustbox}{max width=\columnwidth}
    \footnotesize
    \begin{tabular}{@{} l c c c c @{}}
    \toprule
    \textbf{Reference} & \textbf{Device} & \textbf{LUTs} & \textbf{FFs} & \textbf{Speed} \\
    \midrule
    Kim et al.~\cite{kim}  & Zynq-7030  & 43{,}080 & -- & 1.06\,Gbps \\
    Kang et al.~\cite{8969630} & Zynq-7030  & 2{,}468 & 3{,}379 & 100\,Mbps \\
    Dakhil et al.~\cite{9800025}  & Artix-7  & 119 & 72 & 100\,Mbps \\
    \textbf{Proposed} & \textbf{Virtex-6}  & \textbf{1{,}595} & \textbf{3{,}024} & \textbf{1\,Gbps} \\
    \textbf{Proposed} & \textbf{Versal VCK190}  & \textbf{1{,}348} & \textbf{2{,}655} & \textbf{10\,Gbps} \\
    \bottomrule
    \end{tabular}
    \end{adjustbox}
    \begin{flushleft}
    \scriptsize UART and 10G PCS/PMA excluded for all proposed-design figures. Kim et al.\
    reached the Zynq-7030 BRAM limit; prior-work figures converted from
    datasheets.    
    \end{flushleft}
\end{table}

\subsection{Scaling and Latency}
At the worst-case measured cost of 71 LUTs (Virtex-6) per fingerprint, the Virtex-6 supports approximately 2,100 fingerprints before exhausting LUT resources. Extrapolating the Versal worst case cost ($\approx26$ LUTs and $\approx130$ FFs per fingerprint) suggests capacity for roughly 13,500 fingerprints before flip-flops become limiting, although timing closure has been verified only to 2,000 fingerprints ($f_{\max}=212.5$\,MHz, exceeding the 200\,MHz target; see~\figurename{~\ref{fig:scalability}B}). These estimates are conservative, as the scaling study uses only worst-case mock fingerprints (26 LUTs, 130 FFs each); practical deployments with a mix of simpler and more complex fingerprints are therefore expected to scale at least as well. The operationally relevant metric is throughput (zero drops); the 10\,ns latency follows from the zero-buffer design and leaves headroom for automated responses (e.g.\ RST injection). Software IDS cannot sustain comparable rates: even with infinitely fast regex matching, Snort~3 processes roughly 400\,Mbps per core~\cite{pigasus}, and per-packet latencies are three to four orders of magnitude above the 10\,ns achieved here. 

\subsection{Detection Validation}
\label{sec:detection_validation}
The system was validated in three settings (Table~\ref{tab:validation}). On a passive mirrored departmental uplink, all 396{,}528 TCP-SYN packets in an 8-hour window carried IHL=5/DO=5, of which 113 matched deployed fingerprints. Re-evaluating the same Boolean functions in software on the captured \texttt{.pcap} gave identical results, establishing exact hardware--software equivalence; this verifies implementation correctness. A 500{,}000-packet replay from a network telescope monitoring approximately 65{,}000 unused IPv4 addresses yielded 26{,}555 detections with exact agreement, and 1{,}000 back-to-back fingerprinted packets were all detected at line rate, confirming sustained throughput under worst-case match density.
By design, the system detects only deployed fingerprints; undetected traffic using other patterns cannot be distinguished from benign packets here.

\begin{table}[!t]
\centering
\caption{Detection Validation Results}
\label{tab:validation}
\resizebox{\columnwidth}{!}{%
\begin{tabular}{@{}lrrrr@{}}
\toprule
\textbf{Scenario} & \textbf{Setting} & \textbf{Total pkts} & \textbf{TCP-SYN} & \textbf{Detected} \\
\midrule
Live traffic    & 8\,h mirrored link  & 172{,}800{,}000 & 396{,}528 & 113   \\
Telescope replay & ${\sim}$65K unused IPv4, bulk replay      & 500{,}000     & 500{,}000 & 26{,}555  \\
Stress test     & Back-to-back fingerprinted pkts & 1{,}000       & 1{,}000   & 1{,}000   \\
\bottomrule
\end{tabular}%
}
\end{table}

\section{Conclusions}
This brief presented a lightweight FPGA architecture for real-time, fingerprint-based detection of TCP-SYN port scanning. Performing all detection at the MAC layer and completing before the subsequent frame's header, it achieves line-rate processing with 10\,ns latency, three to four orders of magnitude below software IDS. Each fingerprint is a parallel Boolean function of header fields, so detection latency is independent of signature count while resource cost grows linearly, scaling to over 2,000 fingerprints, with hardware--software equivalence confirmed on production and telescope traffic. The same fingerprint detector is reused across interfaces from GMII/MII (1\,Gbps and below) to XGMII (10\,Gbps), adapting only the physical-interface frontend; the design uses 1.06\% LUTs on a Virtex-6 at 1\,Gbps and 0.5\% on a Versal VCK190 at 10\,Gbps. A Python framework with a defined fingerprint grammar auto-generates the HDL.

\bibliographystyle{IEEEtran}
\bibliography{references}


\end{document}